\documentclass[conference]{IEEEtran}

\IEEEoverridecommandlockouts                              % This command is only needed if 
\usepackage{cite}
\usepackage{amsmath}
\usepackage{graphicx}	% For figure environment
\usepackage{caption}
\usepackage{subfig}
\usepackage{multirow}
\usepackage{stackengine}
\usepackage{amsmath,amssymb,amsfonts}
\usepackage{algorithmic}
\usepackage{textcomp}
\usepackage{xcolor}
\usepackage{hhline}
\usepackage{multirow}
\begin{document}

\title{%Continuously Monitoring Mental Health with Biomarker-Based Prediction of Passage of Time Perception
 %Mental Health  monitoring through wearable Biomarker-Based and continuous Prediction of Passage of time perception.
Wearable and Continuous Prediction of Passage of Time Perception for Monitoring Mental Health
%Mental Health Assessment through Passage of Time Perception Continuously Monitoring based on Wearable Biomarkers
}

%  \author{Lara~Orlandic$^{1}$,~\IEEEmembership{Member,~IEEE,}
%          Adriana~Arza$^{1}$,~\IEEEmembership{Member,~IEEE,}
%          and~David~Atienza$^{1}$,~\IEEEmembership{Fellow,~IEEE}% <-this % stops a space
% \thanks{*This work was not supported by any organization}% <-this % stops a space
% \thanks{$^{1}$L. Orlandic, A. Arza, and D. Atienza are with the Embedded Systems Laboratory (ESL) of
%         the Swiss Federal Institute of Technology (EPFL), 1015 Lausanne, Switzerland.
%         {\tt\small lara.orlandic@epfl.ch}}%
% }

% author names and affiliations
% use a multiple column layout for up to three different
% affiliations

\author{\IEEEauthorblockN{Lara Orlandic, Adriana Arza Valdes, David Atienza}
\IEEEauthorblockA{Embedded Systems Laboratory (ESL), Swiss Federal Institute of Technology Lausanne (EPFL), Switzerland\\
Email: [lara.orlandic, adrana.arza, david.atienza]@epfl.ch}
}

\maketitle
\thispagestyle{empty}
\pagestyle{empty}

%%%%%%%%%%%%%%%%%%%%%%%%%%%%%%%%%%%%%%%%%%%%%%%%%%%%%%%%%%%%%%%%%%%%
\begin{abstract}

A person's passage of time perception (POTP) is strongly linked to their mental state and stress response, and can therefore provide an easily quantifiable means of continuous mental health monitoring. In this work, we develop a custom experiment and Machine Learning (ML) models for predicting POTP from biomarkers acquired from wearable biosensors. We first confirm that individuals experience time passing slower than usual during fear or sadness (p = $0.046$) and faster than usual during cognitive tasks (p = $2 \times 10^{-5}$). Then, we group together the experimental segments associated with fast, slow, and normal POTP, and train a ML model to classify between these states based on a person's biomarkers. The classifier had a weighted average F-1 score of 79\%, with the fast-passing time class having the highest F-1 score of 93\%. Next, we classify each individual's POTP regardless of the task at hand, achieving an F-1 score of 77.1\% when distinguishing time passing faster rather than slower than usual. In the two classifiers, biomarkers derived from the respiration, electrocardiogram, skin conductance, and skin temperature signals contributed most to the classifier output, thus enabling real-time POTP monitoring using noninvasive, wearable biosensors.

\iffalse
\indent \textit{Clinical relevance}— Necessary?
\fi
\end{abstract}


%%%%%%%%%%%%%%%%%%%%%%%%%%%%%%%%%%%%%%%%%%%%%%%%%%%%%%%%%%%%%%%%%%%%%%%%%%%%%%%%
\section{INTRODUCTION}
\bstctlcite{IEEEexample:BSTcontrol}%force et al. if IEEEexample:BSTcontrol has been created in the .bib file (explained in the VII session of the "How to Use the IEEEtran BibTex Style" guide)

The way in which humans perceive the passage of time is a psychological and neurological phenomenon linked to emotions, memory, attention, and the body's response to stress \cite{Droit-Volet2007, VanHedger2017}. When people are busy, amused or excited, they experience time passing faster than it truly does. Conversely, when people are afraid or under stress, time seems to slow down \cite{VanHedger2017, Droit-Volet2016, Droit-Volet2007}. Furthermore, during the recent COVID-19 pandemic lockdown, it was found that the experience of time passing slowly was associated with increased stress, decreased task load, and decreased satisfaction with one's amount of social interactions \cite{OgdenCOVID}. Moreover, a person's passage of time perception (POTP) is a quantifiable measure that is intricately linked to their mental state \cite{Droit-Volet2016}.

Along with distortions in POTP, various emotions also induce changes in physiological processes such as heart rate, blood pressure, muscular contraction and respiration \cite{Valderas2015-HumanRespiration,  Droit-Volet2016}. This phenomenon illustrates the notion that our perception of time is related to our homeostatic state, and thus, to our physiological stress response that our body triggers to deal with a disturbance in homeostatic balance \cite{Lazarus1993}. 

Physiological monitoring through noninvasive biosensors has previously been used as a means of monitoring mental health conditions including depression, anxiety, bipolar disorders, and many more \cite{Garcia-Ceja2016}. These sensors enable continuous measurements and online estimations of a person's emotions \cite{Montesinos2019Multi-ModalDevicesFixed}, thereby providing real-time insights into their mental state and facilitating timely interventions in the case of deteriorating mental health. Biomarkers computed from such sensors have been previously used to classify a person's stress response \cite{Arza2018MeasuringStress, Montesinos2019Multi-ModalDevicesFixed}, cognitive load \cite{Masinelli2020}, and emotions on the arousal-valence scale \cite{Santamaria-Granados2019UsingAMIGOSFixed}. However, no study has thus far has used unobtrusively measured biosignals to predict individuals' POTP, which provides additional insights into their mental state.

\begin{figure*}[ht]
	\centering
	\includegraphics[width=0.88\textwidth]{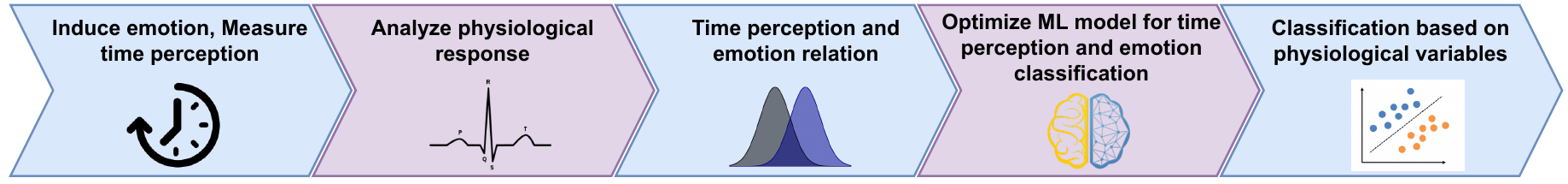}
	\caption{Overall methodology of the POTP statistical analysis and classification}
	\label{Fig:pipeline}
\end{figure*}

In this study, we assess the hypothesis that a person's POTP relates to their stress-related physiological processes, as these relate to their mental state. Thus, we aim to use noninvasive biosignal monitoring and Machine Learning (ML) tools to predict a person's POTP, both directly through binary classification, and indirectly through multi-class prediction of activities that elicit significant changes in POTP. We first investigate the correlations between different induced emotions and their effects on subjects' POTP. Then, we propose a ML model training and optimization procedure for predicting the experimental segment -- and its corresponding POTP -- a subject experienced given their physiological features. Next, we use this ML technique to distinguish periods of fast time perception from those of slow time perception across all experimental segments. Finally, we investigate the influence of individual biomarkers in the POTP and emotion classification outcome to determine which features affect each model.

\section{METHODS}

The set of contributions of this work, namely, the quantification and classification of POTP, is presented in Fig. \ref{Fig:pipeline}. First, various emotional states are induced in the subject through a custom-designed experimental protocol.
Since the physiological stress response has been associated with different levels of cognitive workload, job performance, emotions, and optimal physical states \cite{Arza2018MeasuringStress,Montesinos2019Multi-ModalDevicesFixed}, we investigate multiple biosignals and the correlation of their changes with the perception of passage of time during a variety of stimuli that elicit different emotions, cognitive loads, and stress levels. Thus, we induce emotions and cognitive states on a set of participants as they perform specific tasks and watch emotional short films while we measure their response to these stimuli, as well as their sense of the passage of time.

Throughout the experiment, five biosignals are measured -- electrocardiogram (ECG), skin temperature (SKT), electrodermal activity (EDA), respiration (RSP), and photoplethysmography (PPG) -- for their proven contribution to emotion and psychological stress monitoring  \cite{Montesinos2019Multi-ModalDevicesFixed, Arza2018MeasuringStress}. Next, we develop a ML optimization procedure to select the ML model and hyperparameters that perform the best classification of emotional states and time perceptions based on the subjects' physiological features. Finally, the models are tested to predict subjects' emotional states and time perceptions.

\subsection{Induced Emotional States: Experimental Protocol Design}

In order to investigate the relationship between subjects' emotional state and their POTP, we designed an experiment to elicit various emotions in healthy volunteers using films, cognitive tasks, and intermittent rest states. The exact experimental protocol and the groupings of segments for statistical analysis are displayed in Table \ref{Tab:protocol}:

\begin{table}[ht]
  \centering
  \caption{Experimental protocol}
	\begin{tabular}[t]{l l l l } 
		\hline
		\textbf{No} & \textbf{Segment} & \textbf{Duration} & \textbf{Class} \\ 	\hline
		1 & Relaxation audio    & 3 min     & Rest \\
		2 & Neutral Clip       & 2 min     &  Neutral \\
		3& Rest                & 2 min     & Neutral \\
		4& Fear Clip         & 2 min     & Emotional\\
		5& Mathematics Task    & 3 min     & Cognitive  \\
		6& Rest                & 1.5 min    & Rest \\
		7& Stroop Color Test  & 1.5 min    & Cognitive  \\
        8& Sadness Clip       & 1.5 min    & Emotional\\
        9& Rest                &3 min       & Rest \\
	\hline
	\end{tabular}	
	\label{Tab:protocol}
\end{table}

The video clips were selected from the Emotional Movie \cite{Carvalho2012TheStudy} and the FilmStim Databases \cite{SchaeferAssessingResearchers}, which are two validated databases containing footage to induce specific emotions. The mathematics activity consists of solving arithmetic tasks given time constraints with startling negative feedback \cite{Baumeister2007}. The Stroop Color Test is a color-word reading exercise used to measure cognitive flexibility and working memory \cite{Scarpina2017Fixed}.

Between experimental segments, the subject completes a questionnaire including a visual analogue scale for stress (VASS), in which they select their stress level at that moment from 0 to 100 \cite{Lesage2012ClinicalScale}. Also, the subjects estimate the duration of the past experimental segment using a time scale ranging from 0 to 5 minutes in increments of 30 seconds. Rest phases are included between experimental phases to give the subject time to reset their emotional state to its baseline. During the experiment, the Shimmer Node3 ECG \cite{shimmer3ecgunit} and Empatica E4 wristband \cite{empaticaE4}, which are two lightweight and unobtrusive sensing devices, are used to measure the 5 biosignals.

There were 18 participants recruited for this study: 13 males and 5 females between the ages of 22 and 31. Each experiment was conducted in one sitting using a custom-made Android application to display the instructions and tasks. The ethical approval for this study was obtained from the Cantonal Ethics Commissions for Human Research Vaud (ID 2019-00321).

\begin{figure*} [h!]
    \centering
    \includegraphics[width=0.9\textwidth]{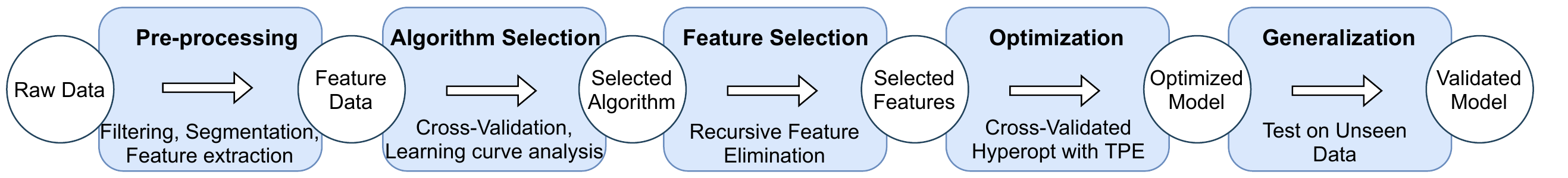}
    \caption{Pipeline for machine learning model selection and testing}
    \label{fig:ml_pipeline}
\end{figure*}
\subsection{Biosignal Measurement and Analysis}

The body's physiological response to emotions and stress is characterized by the combination of several biometric variables \cite{Arza2018MeasuringStress}. These variables can be monitored through parameters derived from biosignals that are unobtrusively measured and continuously monitored. 
After preprocessing each signal (i.e., filtering and delineation), various parameters are extracted. The preprocessing algorithms and primary parameters obtained from each time series signal are computed as in \cite{Montesinos2019Multi-ModalDevicesFixed}, \cite{Masinelli2020}, and \cite{Masinelli2021SPARE}. Next, 80 physiological features in the time and frequency domains are extracted from the parameters in segmentation windows of 45s. Previous works have used windows of length 60s \cite{Montesinos2019Multi-ModalDevicesFixed,DellAgnola2020,Arza2018MeasuringStress}, but learning curves indicated that this produced insufficient data for training. Therefore, we reduced the window length for feature extraction to 45s as a data augmentation technique. These features capture the subject's physiological response during each emotional state. Several key biosignal parameters and features from were the biomarkers are selected, are described as follows: %The key biomarkers selected a that are acquired from several biosignal parameters  are described as follows:

\subsubsection{SKT} 
For the SKT signal, we compute the gradient ($SKT\_gradient$) and total power, ($SKT\_power$) of its power spectral density (PSD).
\subsubsection{EDA} 
The EDA signal is divided into two main components: the skin conductance level (SCL) and the skin conductance response (SCR) as the driver phasic signal ($SCR\_power$). The gradient and mean of the SCL ($SCL\_gradient$, $SCL\_mean$) are obtained.
\subsubsection{RSP} 
We compute respiration rate and period ($RSP\_Rate$,$RSP\_Prd$), duration of air inhaled ($InspTime$) and exhaled ($ExpTime$) and compute statistics based on these parameters. In the frequency domain, we compute the PSD in four different bands of equal bandwidth between 0-1 Hz ($RSP\_PSD_{1-4}$). Additionally, we consider the normalized band power in these four bands ($RSP\_nPSD_{1-4}$), as well as in 5 fine-grained bands in 0.08-0.6 Hz ($RSP\_pBF_{1-5}$). Then, we extract $RSP\_F1pond$, which is the mean frequency of a Gaussian distribution used to fit the PSD estimated in the HF band ($0.15-0.5Hz$).
Moreover, we applied the method proposed in \cite{Hernando2016InclusionAssessment} to compute the estimated respiratory frequency, as the largest peak power ($RSP\_Pk$) of the Lomb-Scargle PSD of respiration using a Welch periodogram. Finally, we compute the average signal power across all windows ($RSP\_power$).
\subsubsection{ECG}
From the ECG, the RR intervals are obtained as in \cite{OrlandicREWARD2019}, and then features are extracted based on the Heart Rate Variability (HRV) analysis \cite{TaskForceo1996}, such as its mean ($ECG\_RR\_mean$), median ($ECG\_RR\_median$) and standard deviation ($ECG\_RR\_SDNN$). Additionally, we compute its normalized bandpower in the very low frequency ($ECG\_RR\_nVLF$), low frequency ($ECG\_RR\_nLF$), and high frequency ($ECG\_RR\_nHF$) bands centered at 0.04 Hz, 0.15 Hz, and 0.4 Hz, respectively.
Non-linear features are also extracted from Poincaré plot indicating vagal and sympathetic function. They are the following: the length of the transverse axis ($ECG\_RR\_T$), the length of the longitudinal axis ($ECG\_RR\_L$), and their ratio, called Cardiac Sympathetic Index ($ECG\_RR\_CSI$), as well as the modified CSI ($ECG\_RR\_CSI\_modified$) \cite{DellAgnola2020}.
\subsubsection{PPG} 
We compute several PPG parameters, including the pulse period ($PPG\_PP$), pulse wave rising time ($PPG\_PRT$), pulse wave decreasing time ($PPG\_PDT$), pulse width until reflected wave ($PPG\_PW$). We then extract ensemble statistics from each parameter, as well as the same frequency analysis as for the RR intervals.

\subsection{Passage of Time Perception Assessment}

In order to quantify subjects' POTP, we define the relative time estimation error metric $t_{rel}$. This metric is calculated based on the correct segment time $t_{correct}$ and the subjects' estimation of the passed time $t_{perceived}$, as shown in Equation \ref{rel_time_err_eq}. A positive $t_{rel}$ means that the person experienced time as passing faster than it truly did, whereas a negative $t_{rel}$ corresponds to the perception of time passing slower. 

\begin{equation}
    t_{rel} = \frac{t_{correct} - t_{perceived}}{t_{correct}}*100
    \label{rel_time_err_eq}
\end{equation}

The segments are grouped into three categories, as displayed in Table \ref{Tab:protocol}: emotional, cognitive, and neutral. Neutral segments are placed after the initial rest period to avoid bias by any previous experimental segment or sensor placement. We then test the statistical significance in the difference of means of $t_{rel}$ for each category to confirm previous hypotheses that each segment corresponds to a given POTP.

\subsection{ML Model of Emotional State and Time Passage} 

Next, we train ML models to predict a person's POTP based on their physiological features. Hence, we perform two classification tasks: a binary classification to determine when each person determines that time is passing faster rather than slower than usual, as well as a multiclass classification of the experimental state of the user (emotional, cognitive, neutral), as these states each correspond to a different POTP.

We compare 8 state-of-the art ML classification algorithms to perform the inference: Logistic Regression (LogReg), Decision Tree Classifier (DTC), k Nearest Neighbor (KNN),
Linear Discriminant Analysis (LDA), Gaussian Naive Bayes (GNB), Support Vector Machines (SVM), Random Forest (RF) and eXtreme Gradient Boosting (XGB). A ML model development pipeline is implemented to ensure generalizability of the chosen model across all subjects, as displayed in Fig. \ref{fig:ml_pipeline}.

First, every biosignal for each experimental segment is divided into 45 s non-overlapping segments. The signals are then preprocessed and the aforementioned features are extracted. Features with multiple NaN values are removed as they may be unstable. Then, all features are scaled by subtracting the mean and dividing by the standard deviation of each feature using the training dataset. In both classification tasks, the same randomly-selected 22\% of subjects is designated as the testing set to assess the generalizability of the trained model. No samples belonging to the same subject appear in both the testing and training sets, nor the training and validation sets of each cross-validation (CV) fold.

Next, we train all eight ML models and perform 10-fold Leave-n-Subjects-Out CV, thereby ensuring that signal segments belonging to the same subject are not used for training and validation at a given CV fold. 20\% of the subjects are used for validation in each fold. The utilized metric is the F-1 score in the case of binary classification, and weighted average F-1 score in the multi-label case. The selected model is the one with the highest mean F-1 score across the training folds. In the case that multiple algorithms produce similar mean F-1 scores, the learning curves of the algorithms are analyzed to examine the bias-variance trade-off. The model with the higher variance than bias is selected, as the subsequent hyperparameter optimization and feature elimination steps intend to reduce overfitting \cite{Munson2009Bagging}. 

Once the algorithm is selected, its hyperparameters are tuned using Tree-structured Parzen Estimators (TPE) \cite{Bergstra2011final} with the objective of maximizing the mean CV F-1 score.
Next, Recursive Feature Elimination with Cross-Validation (RFECV) is performed to remove features that do not contribute to the classification outcome \cite{Munson2009Bagging}. Following RFECV, the aforementioned TPE procedure is performed to re-optimize the model to its new feature set.

\section{RESULTS}

First, we analyze the results of the statistical analysis of relative time errors in the feature set to determine which experimental segments truly produce significant POTP distortions. We use this information, along with the distribution of $t_{rel}$ values, to set up the emotion classification procedure and determine thresholds for the POTP classification. ML models are trained to perform each classification task based on the extracted biomarkers and subsequently analyze each model's generalization capabilities.
\begin{figure} []
    \centering
    \includegraphics[width=0.83\columnwidth]{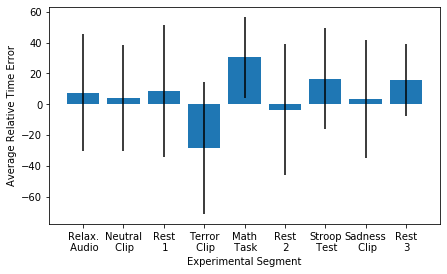}
    \caption{Average $t_{rel}$ (blue), with standard deviations (black), for each experimental segment}
    \label{fig:avg_time_errs}
\end{figure}
\subsection{Passage of Time Perception}
\label{sec:POTP}

The average $t_{rel}$ across all subjects for each segment are displayed in Fig. \ref{fig:avg_time_errs}. We can see that the most positive $t_{rel}$ occurs for the mathematics test, whereas the most negative one occurs for the fear clip.

Next, we performed one-sample, one-tailed t-tests to determine whether or not these time errors deviated significantly from zero during each experimental segment. Tasks corresponding to the same hypothesized POTP direction were grouped together, as described in Section \ref{Tab:protocol}. The null hypothesis for neutral (rest and neutral clip) and emotional (fear and sadness) tasks is that the time error is less than zero, since people typically perceive time as passing slower when they are bored or afraid. Conversely, for the cognitive (mathematics and Stroop) tasks, we ran a right-tailed t-test, since people normally perceive time as passing faster when they are busy. This grouping leads to a nearly balanced sample of neutral, emotional, and cognitive segments.

The results of the t-tests are summarized in Table \ref{pval_table}. The neutral segments showed no significant deviation from zero (p = 0.942). Conversely, the cognitive tasks and the emotional tasks were significantly higher than zero (p = $2.01 \times  10^{-5}$) and lower than zero (p = 0.0456), respectively. The sadness clip showed significant variance in $t_{rel}$ and no significant deviation from zero by itself, perhaps due to the short duration of the clip and subjectivity in individuals' perceptions of sadness.

\begin{table}[ht]
   \centering
   \caption{Statistical Analysis Results}
    \begin{tabular}{|l|l|l|l|}
     \hline
 Class       & Average $t_{rel}$ & P-Value                 & Pass. of Time       \\ \hline
Emotional & -16.1\%                   & 0.0456                  & Slower                \\ \hline
Neutral            & 6.94\%                    & 0.942                   & No change \\ \hline
Cognitive & 23.6\%                   & $2 \times  10^{-5}$ & Faster                \\ \hline
\end{tabular}
\label{pval_table}
\end{table}

The reported VASS stress levels were highest during the mathematics task ($63 \pm 26$) and fear clips ($38 \pm 20$). There was no significant correlation ($\alpha = 0.05$) between the stress levels and either the signed or absolute value of $t_{rel}$.

\begin{figure} []
    \centering
    \includegraphics[width=\columnwidth]{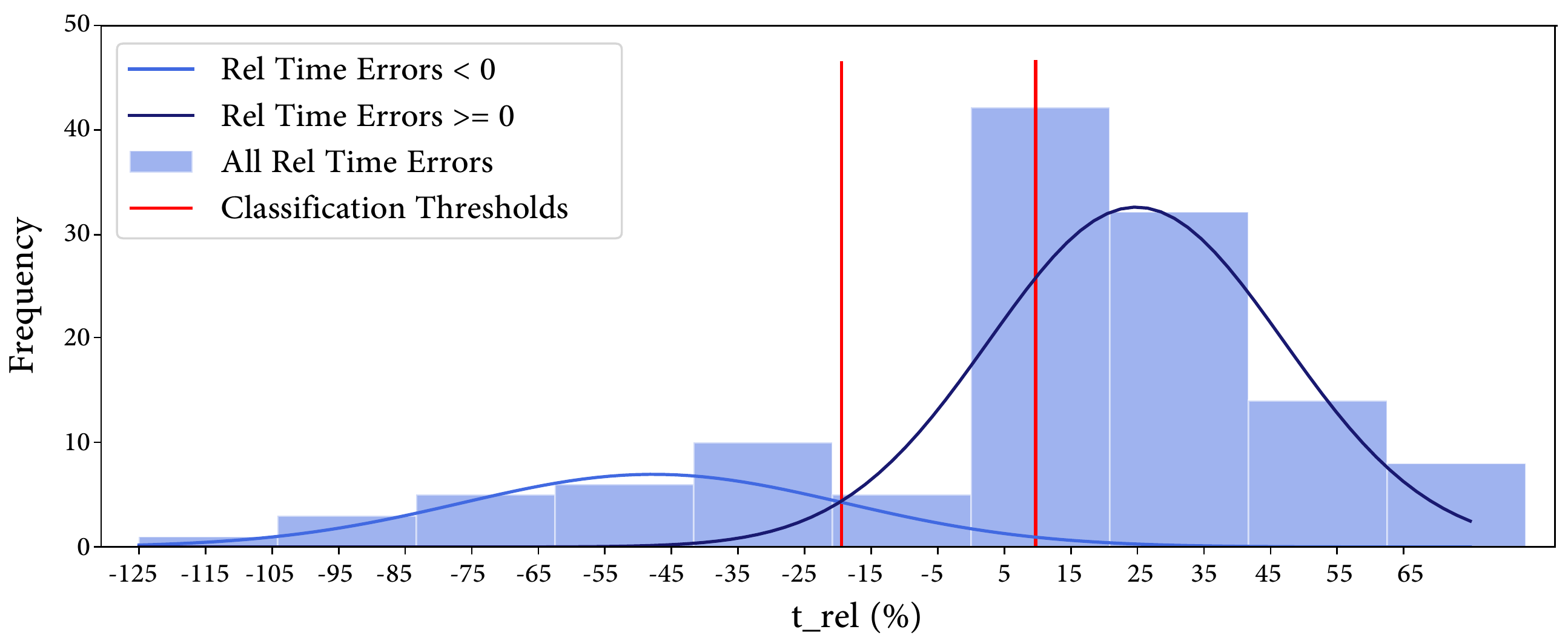}
    \caption{Classification thresholds for $t_{rel}$}
    \label{fig:thresholding}
\end{figure}

Finally, to facilitate classification based directly on POTP, we define thresholds on $t_{rel}$ to identify biosignal segments during which the individual subjects experience time passing significantly fast and slowly. Fig. \ref{fig:thresholding} shows a histogram of all $t_{rel}$ values in the training dataset. We notice a bi-modal distribution of positive and negative time errors, so we fit Gaussian curves to the positive and negative $t_{rel}$ values. The upper threshold is located two standard deviations to the right of the mean of negative $t_{rel}$ values, while the lower threshold is two standard deviations to the left of the positive $t_{rel}$ mean. This process provides an empirical estimation of statistically significant low and high values of $t_{rel}$.

Therefore, our final classification labels for the passage of time are as follows:
\begin{itemize}
    \item Time passes faster (1) if $t_{rel} > 10$
    \item Time passes slower (-1) if $t_{rel} < -19$
\end{itemize}
This grouping leads to an imbalanced sample of the two classes, since there are about twice as many segments in which time passes faster rather than slower.

\subsection{POTP Classification}

\begin{figure} [ht]
    \centering
    \includegraphics[width=0.75\columnwidth]{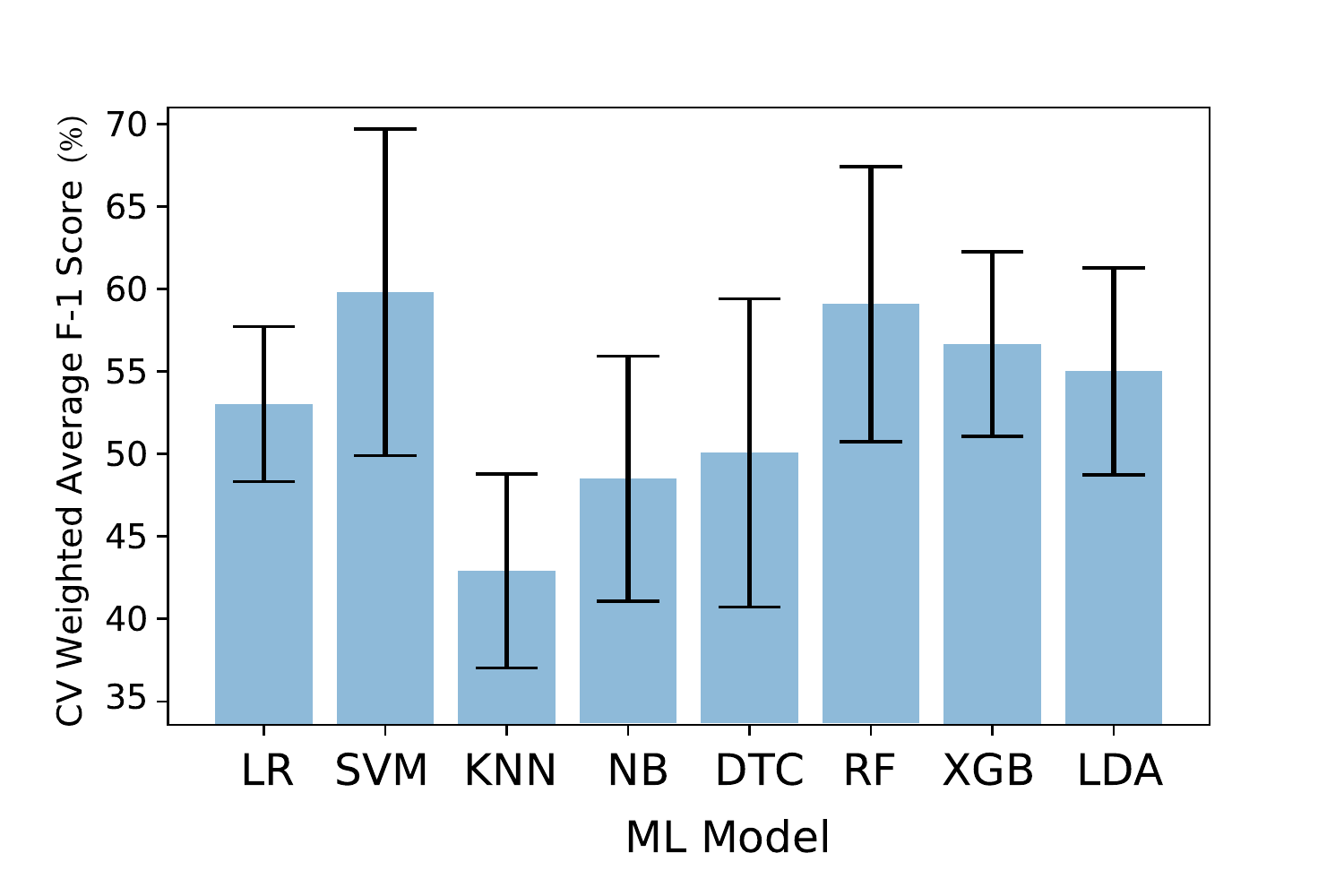}
    \caption{CV mean and st. dev. F-1 scores for each ML model}
    \label{fig:algo_comp_3_class}
\end{figure}

Once we determined which experimental states corresponded to a faster, slower, or normal POTP, we built a ML model to classify these states. We first use the procedure described in Fig. \ref{fig:ml_pipeline} to train a ML model to determine whether a person was in the emotional, cognitive, or neutral phases of the experiment based on their physiological features. The algorithm selection procedure is shown in Fig. \ref{fig:algo_comp_3_class}, which displays the CV F-1 scores of each model.

\begin{figure} []
    \centering
    \includegraphics[width=0.90\columnwidth]{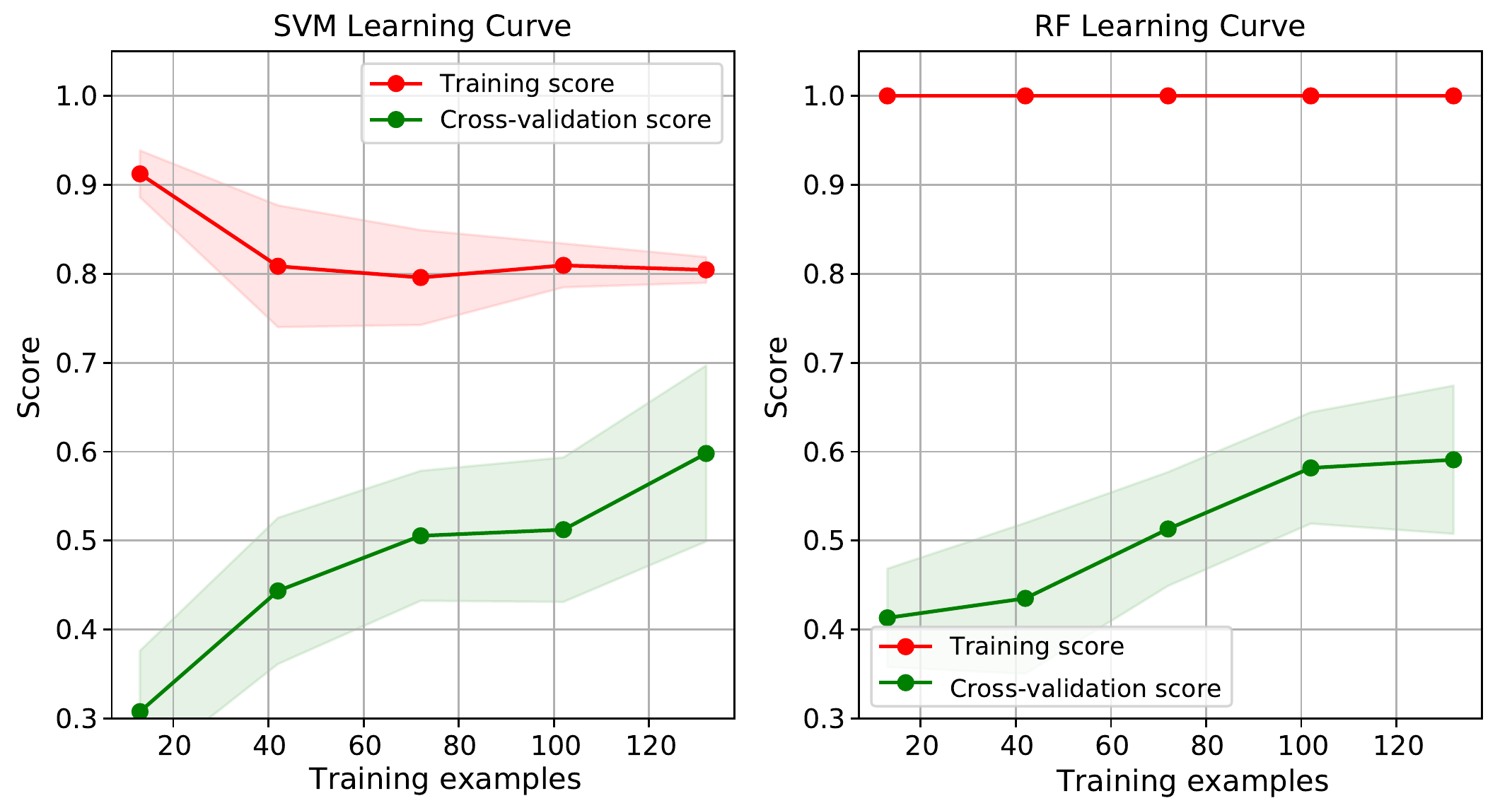}
    \caption{Learning curves of the SVM and RF algorithms}
    \label{fig:learning_curves}
\end{figure}
We note from Fig. \ref{fig:algo_comp_3_class} that SVM exhibits the highest mean F-1 score, but its standard deviation is much larger than that of RF, which has a similar mean. To finalize our model selection, we examine the learning curves, as shown in Fig. \ref{fig:learning_curves}.

We can see that the training and CV accuracies of the SVM model converge quickly to around 0.7, implying low variance and high bias, and in the RF there remains a large gap between the training and validation accuracies, implying high variance. We therefore select the RF model for further analysis because its variance can be reduced using TPE and RFECV. Following TPE, the average CV score of the RF model increased by 7.2\%, as shown in Table \ref{tab:ml_opt_results}. Then, RFECV revealed that the optimal number of features was 45.

\begin{table}[ht!]
   \centering
   \caption{ML Optimization Step-by-Step Results}
    \begin{tabular}{|l|l|l|}
     \hline
 \multirow{2}{*}{ML Step } & \multicolumn{2}{c|}{Mean CV F-1 Score (\%)} \\\hhline{~--}
                        & 3-Class RF &  2-Class XGB\\\hline
First training  &   56.7\%    &69.0\%     \\ \hline
Hyperparameter Opt.    &    63.3\%    & 70.6\%     \\ \hline
RFECV                   &   63.3\%   (45 feat.)  & 70.8\% (18 feat.)   \\ \hline
Hyperparameter Re-Opt.  &   --    & 72.4\%  \\ \hline
Test on Unseen Data               &    79.0\%   & 77.1\%  \\ \hline
\end{tabular}
\label{tab:ml_opt_results}
\end{table}

\begin{figure} []
    \centering
    \includegraphics[width=0.9\columnwidth]{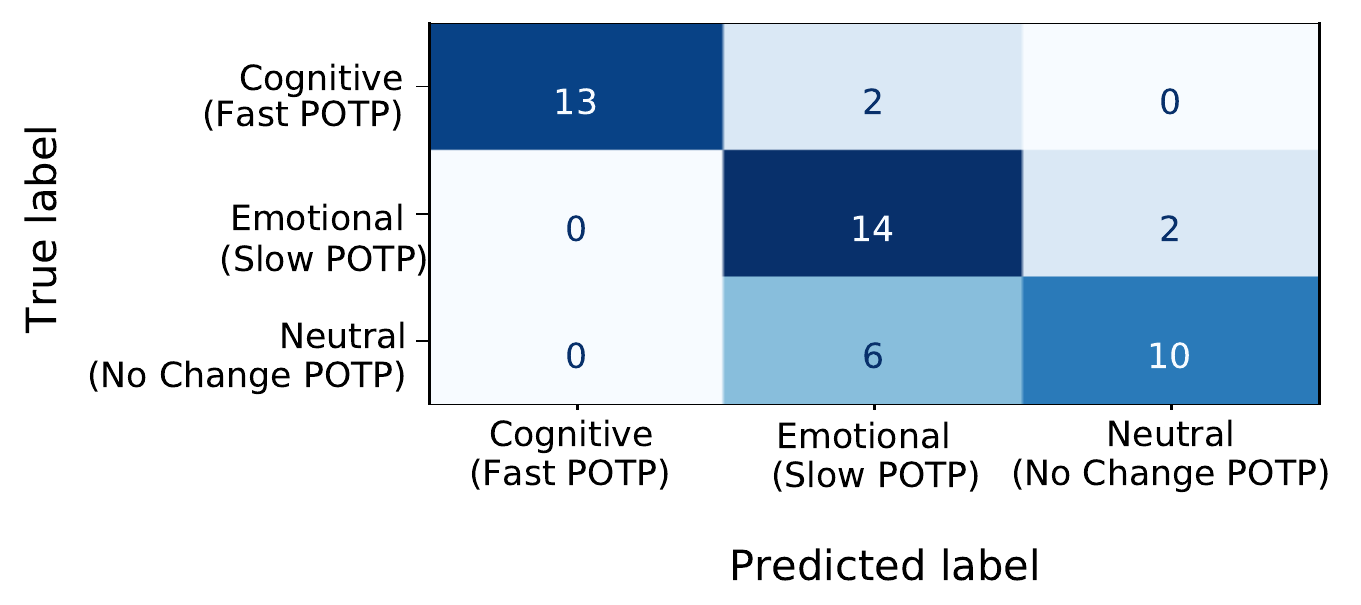}
    \caption{Confusion matrix for the emotion classifier}
    \label{fig:conf_mat_3_class}
\end{figure}

The final model is then tested on new, unseen data from four subjects. The results are displayed in the confusion matrix in Fig. \ref{fig:conf_mat_3_class}, as well as the F-1 scores in Table \ref{tab:emotion_classifier}. We can see that the ``Fast POTP" class is the easiest for the classifier to distinguish, with a 100\% precision and highest F-1 score of 93\%. Most of the misclassifications are due to the ``No Change in POTP" signals being classified as ``Slow POTP" signals. The weighted average of the F-1 scores of all classes with respect to the number of data points per class is 79\%.

\begin{table}[ht]
   \centering
   \caption{Emotion Classifier Results on Unseen data}
    \begin{tabular}{|l|l|l|l|}
     \hline
   Class    & Precision & Recall    & F-1 Score       \\ \hline
Emotional (Slow POTP)   & 64\%     & 88\%         & 74\%               \\ \hline
Neutral  (No change POTP)  & 83\%     & 62\%          & 71\% \\ \hline
Cognitive (Fast POTP) & 100\%     & 87\%          & 93\%                \\ \hline
Weighted Avg. & 82\%     & 79\%          & 79\%                \\ \hline
\end{tabular}
\label{tab:emotion_classifier}
\end{table}

\begin{figure} [hbp!]
    \centering
    \includegraphics[width=\columnwidth]{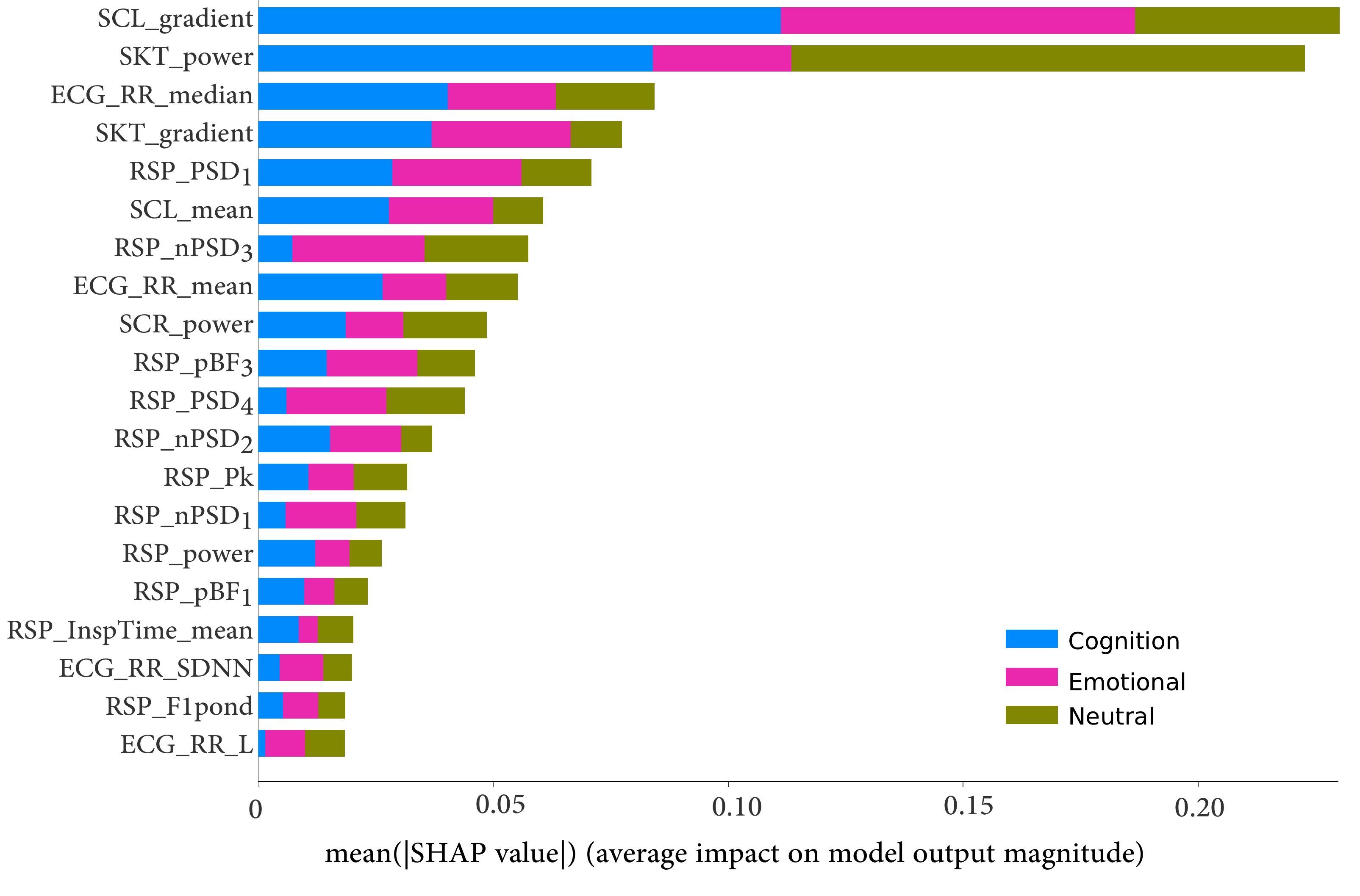}
    \caption{SHAP values indicating the feature importance to the output of the emotion classifier}
    \label{fig:shap_3_class}
\end{figure}

Finally, we compute the Shapley (SHAP) values of the model, which  are measures of the relative importance of the features in the model’s classification decision \cite{Erikstrumbelj}. The features and their relative importances to each class are displayed in Fig. \ref{fig:shap_3_class}. We can see that by far, the most important features are the $SCL\_gradient$ and $SKT\_power$. The remaining features relate to the ECG, RSP, SKT, SCR, and SCL signals.

\begin{figure} [ht]
    \centering
    \includegraphics[width=\columnwidth]{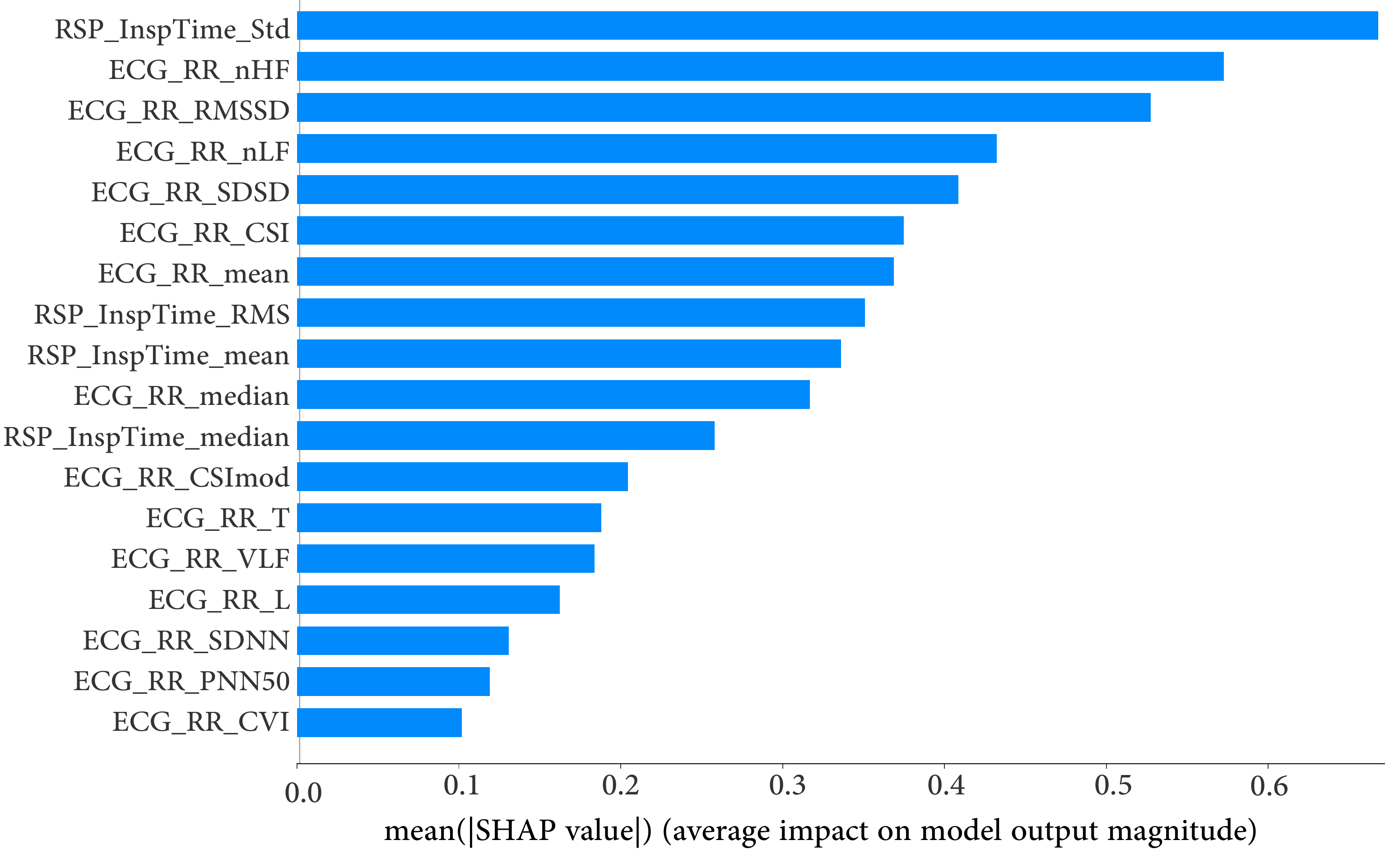}
    \caption{SHAP values of the time perception classifier}
    \label{fig:shap_timelabels}
\end{figure}

The SHAP values of the passage of time perception classifier are displayed in Fig. \ref{fig:shap_timelabels}. In this case, the two most important features are PSD features of the ECG and RSP inspiration time. Other important features are derived from the ECG and RSP signals. All of the important RSP features relate to the time-domain inspiration time parameter, whereas the important ECG features are computed using the time and frequency domain of the R-R interval signal.

\section{DISCUSSION AND FUTURE WORK}

The way a person perceives the passage of time is a quantifiable metric that indicates their mental state. Therefore, continuous monitoring of subjects' POTP may provide insights into their mental well-being. In this work, we have developed an experiment to induce certain emotions that have known effects on the passage of time. Our results have confirmed our hypotheses that people consistently interpret time as passing slower during fear or sadness (p = 0.046), and faster during mentally taxing tasks (p = $2 \times 10^{-5}$). Then, we have developed a classifier to predict which experimental state, which is correlated to a change or lack thereof in POTP, a subject experienced based on physiological features extracted from unobtrusively measured biosignals. This classifier obtained a weighted average F-1 score of 79\%, with fast POTP tasks being the easiest to distinguish with an F-1 score of 93\%.

Next, we classify the POTP directly by identifying segments with significantly high and low $t_{rel}$ values regardless of the task at hand. We obtain a 77.1\% F-1 score in distinguishing time passing fast rather than slow, meaning that is possible to determine a person's POTP based solely on their physiological signals. When we analyzed the feature importance of the two models, the emotion classifier used more diverse biosignals than the POTP classifier, the latter of which did not heavily weigh features derived from SCL or SKT. Both classifiers place heavy importance on the $ECG\_RR\_median$, $ECG\_RR\_mean$, $ECG\_RR\_SDNN$, and $RSP\_InspTime\_mean$ biomarkers. These results indicate that by monitoring a few biosignals with simple, wearable sensors, it is possible to unobtrusively monitor POTP and mental state on a continuous basis.

In the future, these preliminary results may be enhanced by using a larger, more diverse sample of subjects for testing and training, as well as longer experimental segments. These factors may help overcome the high variance seen in $t_{rel}$ of the sadness clip, as well as the class imbalance in the binary classification task. With more individuals and longer biosignal durations, Deep Learning analysis may be employed.

\addtolength{\textheight}{-12cm}   % This command serves to balance the column lengths
                                  % on the last page of the document manually. It shortens
                                  % the textheight of the last page by a suitable amount.
                                  % This command does not take effect until the next page
                                  % so it should come on the page before the last. Make
                                  % sure that you do not shorten the textheight too much.

%%%%%%%%%%%%%%%%%%%%%%%%%%%%%%%%%%%%%%%%%%%%%%%%%%%%%%%%%%%%%%%%%%%%%%%%%%%%%%%%

%%%%%%%%%%%%%%%%%%%%%%%%%%%%%%%%%%%%%%%%%%%%%%%%%%%%%%%%%%%%%%%%%%%%%%%%%%%%%%%%

%%%%%%%%%%%%%%%%%%%%%%%%%%%%%%%%%%%%%%%%%%%%%%%%%%%%%%%%%%%%%%%%%%%%%%%%%%%%%%%%

\section*{ACKNOWLEDGMENT}

This work has been partially supported by the Innosuisse Innocheque (No. 38484.1), the ML-edge Swiss National Science Foundation (NSF) Research project (GA No. 200020182009/ 1), and the MyPreHealth research project (Hasler Foundation project No. 16073). We would also like to thank Giulio Masinelli for his help with the experimental setup.

\bibliographystyle{IEEEtran}
\bibliography{references1.bib, references_stress.bib}

\end{document}